\documentclass[preprint,showpacs,preprintnumbers,amsmath,amssymb]{revtex4}

\usepackage{graphicx}
\usepackage{dcolumn}
\usepackage{bm}

\begin{document}

\title{Rotational Effects of Twisted Light on Atoms Beyond
the Paraxial Approximation\\}

\author{R. J\'auregui}

\affiliation {Instituto de F\'{\i}sica, U.N.A.M., Apdo. Postal
20-364, M\'exico D. F. 01000, M\'exico.\\
e-mail:rocio@fisica.unam.mx\\}

\date{\today}

\begin{abstract}
The transition probability for the emission of a Bessel photon by
an atomic system is calculated within first order perturbation
theory. We derive a closed expression for the electromagnetic
potentials beyond the paraxial approximation that permits a
systematic multipole approximation . The matrix elements between
center of mass and internal states are evaluated for some
specially relevant cases. This permits to clarify the feasibility
of observing the rotational effects of twisted light on atoms
predicted by the calculations. It is shown that the probability
that the internal state of an atom acquires orbital angular
momentum from light is, in general, maximum for an atom located at
the axis of a Bessel mode. For a Gaussian packet, the relevant
parameter is the ratio of the spread of the atomic center of mass
wave packet to the transversal wavelength of the photon.
\end{abstract}

\maketitle
\section{Introduction}
It is important to understand the influence of the angular
momentum of light on the dynamics of atomic systems and
microparticles  from the point of view of basic and applied
physics. The pioneering work by Beth \cite{Beth}  showed that
circularly polarized light has rotational effects on rigid bodies.
Likewise, observed atomic transitions are simply described in
terms of the angular momentum carried by plane wave circularly
polarized photons. Thus, the phenomenological relation between
spin angular momentum and polarization is well established.

In the last few years it has  been shown both theoretically
\cite{Allen} and experimentally \cite{dholakia} that light beams
may carry angular momentum not directly related to their
polarization state. This form of angular momentum is usually
qualified as 'orbital' and is due to an azimuthal phase dependence
of the transverse electromagnetic intensity. For laser beams
properly described within the paraxial approximation, e. g.,
Laguerre-Gaussian beams, the total angular momentum of light can
be clearly divided into spin and orbital parts; this separation
has direct physical consequences in the motion of microparticles
\cite{dholakia,karen}. In the case of atoms, the orbital angular
momentum of paraxial light can induce torques in the center of
mass \cite{babiker3} while the probability of changing the
internal angular momentum is very small \cite{babiker4}. However,
the separation between orbital and spin angular momentum is not so
natural beyond the paraxial approximation \cite{barnett94}. On the
one hand, the general forms of these quantities, as they usually
appear in the literature, are not gauge invariant \cite{de
Broglie, gottfried}, despite the fact that physical observables
are expected to be so. On the other hand, the concept of
polarization for 'twisted' light is not identical to that used for
plane waves since, in general, the electric and/or magnetic field
of twisted beams   are nonzero along the main propagation axis.

Among electromagnetic modes carrying 'orbital' angular momentum,
Bessel modes are particularly  interesting because they propagate
with an intensity pattern invariant along its axis \cite{Durnin}.
Experimental realizations of such beams and their mechanical
effects on microparticles are the subject of many current
investigations \cite{karen}. Studies concerning the separation
between spin and orbit angular momentum have been carried out both
semiclassically \cite{karentesis} and quantum mechanically
\cite{hacyan}. The quantum dynamical properties of the Bessel
photons should, in principle, be studied via the operators
assigned to energy, momentum, orbital angular momentum and spin
using the standard quantum optics formalism. However, direct
calculations show that these operators do not follow the algebra
of the translation and rotation group \cite{hacyan}. Actually, the
standard spin operator has a behavior more similar to an helicity
operator.

The purpose of the present paper is to study the rotational
effects of Bessel photons on atomic systems as a dynamical mean to
measure the mechanical properties of twisted light beyond the
paraxial approximation. To this end, we evaluate the transition
amplitude for the spontaneous emission of a Bessel photon by a non
relativistic hydrogenic atom within first order perturbation
theory. In the next section, we describe the system in detail
including the interaction which is incorporated using minimal
coupling in the Coulomb gauge. Then, an explicit expression
relevant for a systematic multipole approximation is obtained and
it is applied to the explicit calculation of matrix elements of
the interaction Hamiltonian between an unperturbed atom and
electromagnetic field states. Finally, a comparison with previous
results obtained in the paraxial approximation is given, and some
conclusions following our results are summarized.

\section{ The atom-radiation system}
Consider two particles of opposite charges $q_e$ and $q_N$,
gyromagnetic ratios $g_e$ and $g_N$, masses $M_e$ and $M_N$,
 and vector positions ${\bf r}_e$ and
${\bf r}_N$. This hydrogen like atom is assumed to be described to
a good approximation by a non relativistic Hamiltonian of the form
\begin{equation}
\hat H_P = \frac{p_e^2}{2M_e} + \frac{\hat P_N^2}{2M_N} + V_r +
V_R,
\end{equation}
with $V_r$ an internal potential depending on the relative
coordinate ${\bf r} ={\bf r}_e -{\bf r}_N$ and $V_R$ an external
potential that affects the center of mass coordinate ${\bf R}=
(M_e{\bf r}_e + M_N{\bf r}_N)/(M_e +M_N)$. The atom state can be
written as a superposition of wave functions
\begin{equation}
\Psi({\bf r}_e,{\bf r}_N,\chi_e,\chi_N) = \Phi({\bf R}) \phi({\bf
r})\chi_N \chi_ee^{-iEt/\hbar}.
\end{equation}
where
\begin{eqnarray}
\Big[\frac{p_{CM}^2}{2M} + V_R\Big]\Phi({\bf R}) &=& E_{CM}\Phi({\bf R}),\\
\Big[\frac{\hat p^2}{2\mu} + V(r)\Big]\phi({\bf r}) &=&
E_{rel}\phi({\bf r}),\\
 E &=& E_{CM} +E_{rel}.
 \label{eq:separable}
\end{eqnarray}
and $\chi_e$, $\chi_N$ are the spinors  associated to each
particle. In the simplest case
\begin{equation}
V({\bf r}_e -{\bf r}_N) = \frac {q_eq_N}{\vert{\bf r}_e -{\bf
r}_N\vert}
\end{equation}
and $V_R =0$.

We are interested in the perturbative description of the
interaction of the atom with a Bessel mode. Thus, the quantization
of the free radiation field will be done using transverse magnetic
(TM) and transverse electric (TE) Bessel vector potentials; in the
Coulomb gauge, they are given by:
\begin{eqnarray}
{\bf A}^{(TM)}_{\bf\kappa}&=& \frac{c}{2\omega}E_0
e^{i(k_zz-\omega t)}\big[\psi_{m-1}({\bf e}_x +i{\bf
e}_y)-\psi_{m+1}({\bf e}_x
-i{\bf e}_y) \nonumber\\
&-&i\frac{2k_\bot}{k_z} \psi_m{\bf e}_z\big]\label{eq:potentials1}\\
{\bf A}^{(TE)}_{\bf\kappa}&=& \frac{i}{2k_z}E_0 e^{i(k_zz-\omega
t)}\big[\psi_{m-1}({\bf e}_x +i{\bf e}_y)+\psi_{m+1}({\bf e}_x
-i{\bf e}_y)\big]\label{eq:potentials2}
\end{eqnarray}
where ${\bf \kappa}$ denotes the set of quantum numbers
$\{k_\bot,k_z,m\}$,
\begin{equation}
\psi_m(\rho,\phi;k_\bot) = J_m(k_\bot\rho) e^{im\phi},
\end{equation}
$J_m$ is the cylindrical Bessel function of order $m$,
\begin{equation}
 \omega = \sqrt{k_\bot^2 + k_z^2},
 \end{equation}
 and
 \begin{equation}
  E^2_0=\frac{\hbar
  k_\bot}{2\pi}\big[\frac{k_z^2c^2}{\omega^2}\big].
  \label{norm}
\end{equation}
The corresponding electric fields are  ${\bf
E}^{(i)}_{\bf\kappa}=i\omega /c{\bf A}^{(i)}_{\bf\kappa}$ while
the magnetic fields are
\begin{eqnarray}
{\bf B}^{(TM)}_{\bf\kappa}&=& \frac{E_0\omega}{2ck_z}
e^{i(k_zz-\omega t)}\big[\psi_{m-1}({\bf e}_x +i{\bf
e}_y)-\psi_{m+1}({\bf e}_x-i{\bf e}_y)\big]\\
{\bf B}^{(TE)}_{\bf\kappa}&=&\frac{iE_0}{2} e^{i(k_zz-\omega
t)}\big[\psi_{m-1}({\bf e}_x +i{\bf e}_y)+\psi_{m+1}({\bf e}_x
-i{\bf e}_y)
\nonumber\\
&-i&\frac{2k_\bot}{k_z} \psi_m{\bf e}_z\big].
\end{eqnarray}
From the electromagnetic potentials, the operator $\hat {\bf A}$
is obtained as
\begin{eqnarray}
\hat {\bf  A}({\bf r}, t) &=& \sum_{i=TM,TE}
\sum_{m=-\infty}^{\infty} \int_0^{\infty}
 dk_\bot \int_{-\infty}^{\infty} dk_z \big[ \hat a^{(i)}_m(k_z,k_\bot)
{\bf A}^{(i)}_{\bf\kappa}({\bf r}, t) \nonumber \\
&+&\hat a^{(i)\dagger}_m(k_z,k_\bot){\bf A}^{(i)*}_{\bf
\kappa}({\bf r}, t)\big],
\end{eqnarray}
\begin{equation}
\big[\hat a^{(i)}_m(k_z,k_\bot), \hat a^{(i)\dagger}_{m^\prime}
(k_z^\prime,k_\bot^\prime)\big]
=\frac{1}{k_\bot}\delta_{m,m^\prime}
\delta(k_\bot-k_\bot^\prime)\delta(k_z-k_z^\prime).\label{quant}
\end{equation}
The normalization condition imposed on the potentials
(\ref{eq:potentials1}-\ref{eq:potentials2}) guarantees that the
radiation energy operator can be written as \cite{hacyan}
\begin{eqnarray}
\hat H_{R}&=&  \sum_{i,m} \int dk_\bot dk_z ~\hbar\omega\hat{N}_m^{(i)}
,\nonumber\\
\hat{N}_m^{(i)}&=&\frac{1}{2}\Big(\hat a^{(i)\dagger}_m \hat
a^{(i)}_m + a^{(i)}_m \hat a^{(i)\dagger}_m \Big).
\end{eqnarray}
The linear momentum electromagnetic operator is
\begin{equation}
\hat {\bf P}=\hbar \sum_i\int dk_\bot dk_z \big[ k_\bot \hat
\Pi_+^{(i)} ({\bf e}_x-i {\bf e}_y ) +k_\bot \hat \Pi_-^{(i)}
({\bf e}_x + i {\bf e}_y ) +k_z \hat \Pi_3 ^{(i)}{\bf e}_z
\big],\label{linmom}
\end{equation}
where the operators $\hat \Pi^{(i)}_{\pm ,3} (k_\bot , k_z)$ are
\begin{eqnarray}
\hat\Pi_+^{(i)} &=& i \sum_m \hat a^{(i)\dagger}_{m-1}  \hat
a^{{(i)}}_m,
\\
\hat\Pi_-^{(i)} &=&-i \sum_m \hat a^{{(i)}}_{m-1} \hat
a^{(i)\dagger}_m,
\\
\hat\Pi_3^{(i)} &=&\sum_m \hat N^{(i)}_m.
\end{eqnarray}
Another important quantity is the angular momentum
\begin{equation}
{\bf J}=\frac{1}{4\pi c}\int_{\cal V} {\bf r} \times \big[ {\bf
E}({\bf r},t) \times  {\bf B}({\bf r},t)\big]dV  .
\end{equation}
Taking ${\cal V}$ as the whole space, using the standard
decomposition:
\begin{eqnarray}
{\bf J}&=& \frac{1}{4\pi c} \int_{\cal V}
 E_i[{\bf r}  \times \nabla]  A_i  dV
+\frac{1}{4\pi
c} \int_{\cal V} ~ {\bf E}\times  {\bf A} ~dV  \nonumber \\
&-&\frac{1}{4\pi c} \oint_{\cal S} {\bf E}\big[{\bf r}  \times
{\bf A}\big]\cdot d{\bf s}, \label{eq:angmom}
\end{eqnarray}
the expression
\begin{eqnarray}
\hat L_z({\bf 0})&=&\frac{1}{4\pi c} \int_{\cal V}   ~
 \hat E_i[{\bf r}  \times \nabla]_z  \hat A_i  ~dV\nonumber\\
 &=&\hbar \sum_{i,m}\int dk_\bot dk_z m
\hat{N}_m^{(i)}
\end{eqnarray}
follows for the orbital angular momentum along the $z$ axis and,
\begin{eqnarray}
\hat W_z&=&\frac{1}{4\pi c} \int_{\cal V} ~ ({\bf E}\times  {\bf
A})_z
~dV \nonumber\\
&=&\hbar \sum_m \int dk_\bot dk_z \frac{c}{2\omega}ik_z \big(\hat
a^{(TM)\dagger}_m  \hat a^{(TE)}_m - \hat a^{(TM)}_m \hat
a^{(TE)\dagger}_m\big)\label{eq:spin}
\end{eqnarray}
represents the helicity operator. The surface integral in
Eq.~(\ref{eq:angmom}) is not well defined for the elementary $TE$
and $TM$ modes because they do not decay rapidly enough as
$r\rightarrow \infty$.

 The Hamiltonian describing the system formed by the atom and the electromagnetic
 radiation is taken to be
\begin{equation}
\hat H = \hat H_P + \hat H_R + \hat H_I,
\end{equation}
with $\hat H_I$ the interaction Hamiltonian. The latter results
from minimal coupling of the particles and the electromagnetic
field in Coulomb gauge, as well as the magnetic interaction
between the magnetic moment $g_i{q_i}/{2M_i}{\bf S}_i$ associated
to the spin of each particle ${\bf S}_i$ with the radiation
magnetic field ${\bf B}$:
\begin{eqnarray}
\hat H_I &=& \hat H_{I1} + \hat H_{I2} + \hat H_{I3}\\
\hat H_{I1} &=& -\sum_{i=1}^2 \frac{q_i}{M_i} {\bf p}_i\cdot
\hat{\bf
A}({\bf r}_i)\\
\hat H_{I2} &=& \sum_{i=1}^2 \frac{q_i^2}{2M_i} \vert \hat{\bf A}({\bf r}_i)\vert^2\\
\hat H_{I3} &=& -\sum_{i=1}^2g_i\frac{q_i}{2M_i}{\bf S}_i\cdot
\hat{\bf B}({\bf r}_i).
\end{eqnarray}

\section{Matrix elements of the interaction Hamiltonian}

\subsection{ The interaction Hamiltonian $H_{I1}$.}
The first order perturbation theory probability amplitude of
emission of a Bessel photon ${\bf A}_{\bf\kappa}^{(i)}$ when the
atom makes a transition between an initial $\Psi_0$ and a final
state $\Psi_F$ via the interaction Hamiltonian $H_{I1}$ can be
written as
\begin{eqnarray}
\langle F,1_\kappa^{(i)}\vert H_{I1}\vert 0;0\rangle &=&
\frac{1}{i\hbar}(E^{(0)}_{CM} - E^{(F)}_{CM})\int d^3rd^3R
\big[\Psi^*_F({\bf r},{\bf R}){\bf R}\Psi^*_0({\bf r},{\bf
R})\big]\nonumber
\\
&\cdot& \big[q_e{\bf A}_{\bf \kappa}^{(i)*}({\bf
R}+\frac{\mu}{M_e}{\bf r}) +q_N{\bf A}^{(i)*}_{\bf \kappa}({\bf
R}-\frac{\mu}{M_N}{\bf r})\big]\nonumber
\\
&+&\frac{1}{i\hbar}(E^{(0)}_{rel} - E^{(F)}_{rel})\int d^3rd^3R
\big[\Psi^*_F({\bf r},{\bf R}){\bf r}\Psi^*_0({\bf r},{\bf
R})\big]\nonumber \\
&\cdot& \big[q_e\frac{\mu}{M_e}{\bf A}_{\bf \kappa}^{(i)*}({\bf
R}+\frac{\mu}{M_e}{\bf r}) -q_N\frac{\mu}{M_N} {\bf A}_{\bf
\kappa}^{(i)*}({\bf R}-\frac{\mu}{M_N}{\bf
r})\big]\label{eq:trans}
\end{eqnarray}
as long as the separability conditions (\ref{eq:separable}) are
satisfied.

Let us define
\begin{equation}
\xi_{lm}(\rho,\varphi;k_\bot) = J_l(k_\bot\rho) e^{im\varphi}
\end{equation}
and consider the case where the argument of the Bessel function
refers to a transverse vector $\vec \rho$ that can be written as
the vector sum of two transverse vectors $\vec \rho $ = ${\bf
R}_\bot -{\bf q}_\bot$. For $l> 0$ the Gegenbauer sum rule
\cite{abramowitz} establishes that
\begin{eqnarray}
\frac{J_l(k_\bot\rho)}{(k_\bot \rho)^l}&=&
2^l(l-1)!\sum_{v=0}^{\infty}(l+v)\frac{J_{l+v}(k_\bot
R_\bot)}{(k_\bot R_\bot)^l}\frac{J_{l+v}(k_\bot q_\bot)}{(k_\bot
q_\bot)^l}C_v^l(\cos(\varphi_R-\varphi_q))\nonumber \\
C_v^l &=& \sum_{s=0}^v
\frac{\Gamma(l+s)\Gamma(l+v-s)}{s!(v-s)!(\Gamma(l))^2}
\cos((v-2s)(\varphi_R-\varphi_q)),\label{eq:multibessel}
\end{eqnarray}
while it can be shown that \cite{babiker4}
\begin{equation}
e^{im\varphi} = \sum_{n=0}^m (-1)^n \left({\begin{array}{*{20}c}
   m  \\n\end{array}}\right)
e^{i(m-n)\varphi_R}e^{in\varphi_q} \Big( \frac{(k_\bot
R_\bot)^{m-n}(k_\bot
q_\bot)^n}{(k_\bot\rho)^m}\Big).\label{eq:multiphase}
\end{equation}
Thus, for $m > 0$
\begin{eqnarray}
\psi_m(\rho,\varphi;k_\bot)  &=&\xi_{mm}(\rho,\varphi;k_\bot)
\nonumber\\&=& 2^m(m-1)! \sum_{v=0}^\infty (m-v)
\frac{J_{m+v}(k_\bot
R_\bot)J_{m+v}(k_\bot q_\bot)}{(k_\bot q_\bot)^m} \nonumber\\
&\cdot&\sum_{s=0}^v
\frac{\Gamma(m+s)\Gamma(m+v-s)}{s!(v-s)!(\Gamma(m))^2}
\cos((v-2s)(\varphi_R-\varphi_q))\nonumber\\
&\cdot&\sum_{n=0}^m (-1)^n\left({\begin{array}{*{20}c}
   m  \\n\end{array}}\right)
\Big(\frac{q_\bot}{R_\bot}\Big)^n
e^{i(m-n)\varphi_R}e^{in\varphi_q},\label{eq:multipolar}
\end{eqnarray}
while, for $m= 0$  \cite{abramowitz}
\begin{equation}
\psi_0(\rho,\varphi;k_\bot) =J_0(k_\bot\rho) =
\sum_{v=-\infty}^\infty J_v(k_\bot R_\bot)J_v(k_\bot
q_\bot)\cos(v(\varphi_R -\varphi_q)).\label{eq:multipolar2}
\end{equation}

 These equations are the basis of the multipolar
expansion of the transition amplitude (\ref{eq:trans}) in
cylindrical coordinates. The relevant values of the vector ${\bf
q}$ are ${\bf q}=(\mu/M_e) {\bf r}$ and ${\bf q} =- (\mu /M_N)
{\bf r}$. Before performing explicit multipole calculations, we
show an important result for the emission of a Bessel photon valid
when the center of mass and relative wave functions are of the
form
\begin{eqnarray}
\Phi({\bf R}) &=&\frac{1}{\sqrt{2\pi}}
e^{im_R\varphi_R}\Upsilon_{CM}(R_\bot,z_R)\nonumber
\\
\phi({\bf r}) &=& \Theta(r)
Y_{l_rm_r}(\theta,\varphi_r)\label{eq:cmrelwf}
\end{eqnarray}
with $Y_{lm}$ the spherical harmonics. In this case, the
integration over the azimuthal angles $\varphi_R$ and $\varphi_r$
leads to selection rules related to the conservation of angular
momentum in the $z$ direction. A direct examination of
Eq.~(\ref{eq:trans}) and
Eqs.~(\ref{eq:multipolar}-\ref{eq:multipolar2}) shows that these
selection rules are of the form
\begin{eqnarray}
m-n\pm v\mp 2s-m&=&m_{R}-m_{R}^\prime \nonumber\\
n\mp v\pm 2s&=&m_{r}-m_{r}^\prime \label{recoil:1}
\end{eqnarray}
for the transition amplitudes proportional to $E^{(0)}_{CM} -
E^{(F)}_{CM}$,  and
\begin{eqnarray}
m-i-n\pm v\mp 2s&=&m_R-m_R^\prime\nonumber\\
i+ n\mp v\pm 2s&=&m_r-m_r^\prime \label{recoil:2}
\end{eqnarray}
with $i=\pm 1,0$, for the transition amplitudes proportional to
$E^{(0)}_{rel} - E^{(F)}_{rel}$. Here the letters $n$, $s$ and $v$
denote the summation indices as they appear in
Eq.(\ref{eq:multipolar}). The first (second) equality in Eqs.~
(\ref{recoil:1}-\ref{recoil:2}) permits the identification of the
angular momentum acquired by the center of mass (internal motion)
in the corresponding emission process. Notice that the total
change in the projection of the angular momentum of the atom along
the $z$ axis {\bf is always $-m\hbar$}. In order to make this
result compatible with the conservation of angular momentum, it is
necessary to consider that the $total$ angular momentum of the
Bessel photon is $m\hbar$. Thus the helicity term in equation
(\ref{eq:angmom}) obtained from standard quantum optics
definitions must be somehow compensated by the surface integral.

Now, consider the cases for which the longitudinal and transverse
long wavelength approximations  $k_z z_r\ll 1$ and $k_\bot \rho_r
\ll 1$ are valid. These conditions are satisfied for paraxial
optical Bessel beams and standard atomic systems.  Notice that
$0\le n \le m$ is the index related to a series expansion on
powers of $q_\bot /R_\bot$ in Eq.~(\ref{eq:multipolar}), and that
due to the relation
\begin{equation}
\frac{J_{m+v}(k_\bot q_\bot)}{(k_\bot q_\bot)^m}=
\Big(\frac{k_\bot q_\bot}{2}\Big)^v \sum_{t=0}^\infty
(-1)^t\frac{(k_\bot q_\bot)^{2t}}{2^{2t}t!(m+v+t)!},
\end{equation}
$v$ can be regarded as an index related to a power expansion
useful for a long wavelength approximation. For $k_\bot \rho_r \ll
1$ the term $v=0$ is expected to be dominant in the series
expansion of the vector potential and the functions $\psi_m$ can
be approximated by
\begin{equation}
\psi_m\sim e^{ik_zz_R-i\omega t}J_m(k_\bot R_\bot)\sum_{n=0}^m
(-1)^n \left({\begin{array}{*{20}c}
   m  \\n\end{array}}\right)
\Big(\frac{q_\bot}{R_\bot}\Big)^n
e^{i(m-n)\varphi_R}e^{in\varphi_q}.\label{eq:centered_cm}
\end{equation}
If the atom is located outside the axes of the Bessel beam then,
in general, $q_\bot \ll R_\bot$ and the $n=0$ term is dominant.
Under such  conditions the neutral atom ($q_e=-q_N$) transition
amplitude
\begin{eqnarray}
\langle F,1_K^{(i)}\vert H_{I1}\vert
0;0\rangle&\sim&\frac{q_e}{i\hbar}(E_{rel}^0 - E_{rel}^F) \int
d^3R\Phi^*_0({\bf R}){\bf A}_K^{(i)*}({\bf R})\Phi_F({\bf
R})\nonumber\\
&\cdot& \int d^3r\phi_F^*({\bf r}){\bf r}\phi_0({\bf
r}).\label{eq:dipole}
\end{eqnarray}
contains the standard dipole matrix element for the relative
coordinates. If the center of mass and internal wave functions are
 given by Eq.~(\ref {eq:cmrelwf}), the transition
amplitude can be written as
$$ \langle F,1_K^{TE}\vert H_{I1}\vert
0;0\rangle\sim \frac{q_eE_0}{k_z\hbar}(E_{rel}^0 - E_{rel}^F)
e^{-i(\omega t-(E_F-E_0)t/\hbar)} \nonumber
$$
\begin{equation}
\cdot\sum_{j=\pm 1}
\delta_{m-j,m_R-m_R^\prime}\delta_{j,m_r^\prime-m_r}
I_{CM}^{(0)}(k_\bot,k_z,m-j)I_{rel}(k_\bot,k_z,j),
\label{eq:recoildipolar1}
\end{equation}
while
$$ \langle
F,1_K^{TM}\vert H_{I1}\vert 0;0\rangle\sim
\frac{q_eE_0c}{i\hbar\omega}(E_{rel}^0 - E_{rel}^F) e^{-i(\omega
t-(E_F-E_0)t/\hbar)}, \nonumber
$$
$$
\cdot\Big[-\frac{2k_\bot
i}{k_z}\big(\delta_{m,m_R^\prime-m_R}\delta_{m_r^\prime,m_r}
I_{CM}^{(0)}(k_\bot,k_z,m)I_{rel}(k_\bot,k_z,0)\nonumber $$
\begin{equation}
+ \sum_{j=\pm 1}(-j)
\delta_{m-j,m_R-m_R^\prime}\delta_{j,m_r^\prime-m_r}
I_{CM}^{(0)}(k_\bot,k_z,m-j)I_{rel}(k_\bot,k_z,j)\big)\Big],
\label{eq:recoildipolar2}
\end{equation}
with
 \begin{equation}
I_{CM}^{(0)}(k_\bot,k_z,m-j)=\int dR_\bot dz \Upsilon_{CM}^{F
*}(R_\bot,z)e^{ik_zz}J_{m-j}(k_\bot R_\bot)\Upsilon_{CM}^{0
}(R_\bot,z),
\end{equation}
and
$$
I_{rel}(k_\bot,k_z,j)=\frac{1}{2l^\prime_r +1}\Big[ \delta_{j,0}
[(l_r^\prime -\vert m_r\vert +1)\delta_{l_r,l^\prime_r
+1}+(l_r^\prime -\vert m_r\vert -1)]\nonumber
$$
\begin{equation}
+(1-\delta_{j,0})[\delta_{l_r,l^\prime_r +1}\delta_{\vert
m_r\vert,\vert m^\prime_r\vert -1}-\delta_{l_r,l^\prime_r
-1}\delta_{\vert m_r\vert,\vert m^\prime_r\vert +1}]\Big] \int dr
r^3\Theta_F^*(r)\Theta_0(r).
\end{equation}

As expected, the selection rules for transitions involving the
relative motion of the charged particles are the same as those
obtained with a plane wave expansion of the radiation potential.
Notice, however, that according to
Eqs.~(\ref{eq:recoildipolar1}-\ref{eq:recoildipolar2}), the
emission of a Bessel photon of orbital angular momentum $m\hbar$
yielding $m_r^\prime = m_r \pm 1$ leads to a rotational recoil
effect $m_R^\prime = m_R -m \mp 1$ for the center of mass while,
to this order of approximation, transitions with $m_r^\prime=m_r$
and $m_R^\prime = m_R -m $ are allowed just for the emission of
transverse magnetic photons. These transitions favor torque
effects on the center of mass and are relevant in the emission of
non paraxial photons.

Going beyond the approximation corresponding to
Eq.~(\ref{eq:dipole}) requires to consider both terms with $n> 0$
in the series (\ref{eq:centered_cm}) and terms with $v>0$ in Eqs.
(\ref{eq:multipolar}-\ref{eq:multipolar2}). In the case of atoms
with $M_N>>M_e$,  the energy involved in changes of the center of
mass motion is usually several orders of magnitude smaller than
the energy involved in changes of the internal state. As a
consequence, the important terms are those proportional to
$E^{(0)}_{rel} - E^{(F)}_{rel}$ in Eq.~(\ref{eq:trans}); for them,
to lowest order in $\mu/M_N$ and first order in $k_\bot r_\bot$:
$$\big[\frac{\mu}{M_e}
\xi_{mm}({\bf R}+\frac{\mu}{M_e}{\bf r}) -\frac{\mu}{M_N}
\xi_{mm}({\bf R}-\frac{\mu}{M_N}{\bf r})\big]\nonumber $$

$$ \sim
e^{im\varphi_R}J_m(k_\bot R_\bot) + \big(\frac{\mu}{M_e}\big)^2
k_\bot r_\bot J_{m+1}(k_\bot
R_\bot)\cos(\varphi_R-\varphi_r)e^{im\varphi_R}\nonumber\\
$$
\begin{eqnarray}
&+&(1-\delta_{m,0})\frac{\mu}{M_e}\Big[\sum_{n=1}^m
\left({\begin{array}{*{20}c}
   m  \\n\end{array}}\right)
 \Big( \frac{r_\bot}{R_\bot}\Big)^n
e^{i(m-n)\varphi_R}e^{in\varphi_r}\Big]\nonumber \\
&\cdot&
 [J_m(k_\bot R_\bot) +\frac {\mu}{M_e } J_{m+1}(k_\bot
R_\bot)k_\bot r_\bot \cos (\varphi_R-\varphi_r)
],\label{eq:quadrupole}
\end{eqnarray}
and at the same order in $k_zq_z$,
\begin{equation}
e^{ik_z(z_R+q_z)}\sim e^{ik_zz_R}(1+ik_zq_z).
\end{equation}
 When inserted in the
transition probability amplitude, Eq.~(\ref{eq:trans}), the terms
proportional to $r_\bot$  and $z$ in the last two equations lead
to matrix elements of the relative motion of the electric
quadrupole type $x_ix_j$. As a consequence, standard selection
rules are obtained for that degree of freedom, $e.$ $g.$, $\Delta
m_z^{rel}=\pm 2\hbar,\pm 1\hbar,0$.

The terms in Eq.~(\ref{eq:quadrupole}) with $n>0$ are expected to
be relevant only when the atomic center of mass is located close
to the axis of symmetry of the Bessel mode. However, there is a
vortex of charge $m$ at that axis and $J_m(k_\bot\rho) = 0$ for
$\rho=0$ if $m\ne 0$. Thus, even if the center of mass is properly
located, the matrix element is expected to be small. To explicitly
quantify this effect, consider two specific situations. In the
first, the center of mass states correspond to a free atom
\begin{equation}
\Upsilon_{CM}^{Free}(R_\bot,z_R) = N_{CM}J_{m_R}(k_\bot^{R}R_\bot)
e^{ik_z^{R}z_R}.
\end{equation}
In the second, the atom is trapped by an external  harmonic
potential
\begin{equation}
\Upsilon_{CM}^{H.O.}(R_\bot,z_R) = N_{CM}e^{-
R_\bot^2/2\alpha^2}\big(\frac{ R_\bot}{\alpha}\big)^{\vert
m_R\vert} L^{\vert m_R\vert}_{\bar n}( R_\bot^2/\alpha^2)
e^{ik_z^{R}z_R}.
\end{equation}
here $\bar n=(N_R-m_R)/2$, and  $N_R$ is the  quantum number
giving the energy of the oscillator $E_R=\hbar\Omega_R (N_R+ 1)$,
$\Omega_R$ is the frequency of the oscillator and $\alpha =
\sqrt{\hbar/M_N\Omega_R}$ is the natural amplitude of the
oscillator.

For simplicity, in both the free and trapped atom cases,  we take
the symmetry axis of the center of mass motion to coincide with
the axis of the Bessel mode. According to Ref.~\cite{gradshtein},
if
\begin{equation}
{\cal I}(k_\bot,k_\bot^R,k_\bot^{R\prime},m,n)=:\int_0^\infty
J_m(k_\bot R_\bot)R_\bot^{-n+1}J_{m_R}(k_\bot^R
R_\bot)J_{m_R+m-n}(k_\bot^{R\prime}R_\bot)
\end{equation}
then \begin{equation} {\cal
I}(k_\bot,k_\bot^R,k_\bot^{R\prime},m,n) =0,\label{eq:free1}
\end{equation}
whenever $k_\bot^{R\prime}>k_\bot +k_\bot^R$, while
$$
{\cal I}(k_\bot,k_\bot^R,k_\bot^{R\prime},m,n)=2^{-n+2}k_\bot^m
(k_\bot^R)^{m_R} (k_\bot^{R\prime})^{m+m_R-2}
\frac{\Gamma(3(m_R+m-n)/2+2)}{m!m_R!}\cdot \nonumber
$$
\begin{equation}
\sum_{u,v=0}^\infty
\frac{(m_R+m-n+1)_{u+v}}{(1+m)_u(1+m_R)_vu!v!}\Big(\frac{k_\bot}{k_\bot^{R\prime}}\Big)^{2u}
\Big(\frac{k_\bot^R}{k_\bot^{R\prime}}\Big)^{2v}\label{eq:free2}
\end{equation}
otherwise.

Equation (\ref{eq:free1}) reflects the conservation of transverse
momentum, and equation (\ref{eq:free2}) shows that the transition
amplitude for $n>0$ decreases as $n$ increases (each coefficient
in the series expansion of positive terms decreases as $n$
increases). Notice also that a free center of mass wave function
is normalized in terms of delta distributions. For free atoms, a
comparison with experimental results would require working with
wave packets.

For an atom trapped in a harmonic potential, a direct use of the
integral
$$
\int_0^\infty dx x^{\nu +1} e^{- x^2/\alpha^2} L_\lambda^{\nu
-\sigma}(x^2/\alpha^2)L_\eta^{\sigma}( x^2/\alpha^2) J_\nu(kx)
\nonumber
$$
\begin{equation} =(-1)^{\lambda+\eta}(2/\sqrt{\alpha})^{-\nu-1}k^\nu
e^{-\frac{\alpha^2k^2}{4}}L^{\sigma-\lambda-\eta}_\eta(\alpha^2k^2/4)
L_\lambda^{\nu-\sigma+\lambda-\eta}(\alpha^2k^2/4)
\end{equation}
shows that even for $n=0$ the transition amplitude depends
exponentially on the ratio of the spread of the harmonic wave
function and the transversal wavelength $k_\bot^2\alpha^2$. The
case $n>0$ can be treated analytically if the center of mass
initial wave function corresponds to the ground state oscillations
($N=0$ and $m_R=0$). In that case, we can use the expression
\cite{prudnikov}:
$$
\int_0^\infty dx x^{\nu -1}J_\mu(b\sqrt{x}) L_n^\lambda(cx)
e^{-cx}=\frac{b^\mu(1-\nu-\mu+\lambda)_n}{2^\mu
n!c^{\nu+\mu/2}}\cdot\nonumber $$
\begin{equation}
\cdot\frac{\Gamma(\nu+\mu/2)}{\Gamma(\mu+1)}
\cdot_2F_2(\nu+\mu/2,\nu+\mu/2-\lambda;\nu+\mu/2-\lambda-n,\mu+1;-\frac{b^2}{4c})
\end{equation}
to show that
\begin{eqnarray}
{\cal J}(\bar n,\alpha,k_\bot,m)
&=&(\sqrt{\alpha})^{n-m}\int_0^\infty
R_\bot^{m-2n+1}e^{\rho_\bot^2/\alpha^2}J_{m}(k_\bot
R_\bot)L_{\bar n}^{m-n}(R_\bot^2/\alpha^2)\nonumber \\
&=& \frac{k_\bot^{m+2n}}{2^{m+2n+1}\bar n!(\sqrt{\alpha})^{-\bar n
-1}}\sum_{r=0}^\infty\frac{(m+\bar n-n+r)!}{(m+\bar
n+r)!r!}\big(\frac{-k_\bot^2\alpha^2}{4}\big)^r.\label{eq:ho2}
\end{eqnarray}
Notice that the terms of this series with $r\gg n$ behave as the
exponential series terms, so that series (\ref{eq:ho2}) is
convergent even if $k_\bot^2\alpha^2/4 > 1$.

\subsection{Matrix elements of the interaction term $H_{I2}$.}

The lowest order contribution of the interaction $\hat H_{I2}$
  to spontaneous emission of
light from vacuum is quadratic in the coupling constant $q_i$ and,
to this order,  involves necessarily two photons. For a hydrogenic
atom, the value of the electron-proton mass ratio implies that
$q_e^2/2M_e\gg q_N^2/2M_N$ and also:
\begin{equation}
\hat H_{I2} \sim \frac{q_e^2}{2M_e} \vert \hat{\bf A}\big({\bf
R}+\frac{\mu}{M_e}{\bf r}\big) \vert^2.
 \end{equation}
In the long wavelength limit, one can use the approximation to the
products
\begin{equation}
\psi_m({\bf R}_\bot +\frac{\mu}{M_e}{\bf
r}_\bot,z)\psi_{m^\prime}({\bf R}_\bot +\frac{\mu}{M_e}{\bf
r}_\bot,z)\sim e^{i(k_z+k_z^\prime)z_R-i(\omega+\omega^\prime)
t}J_m(k_\bot R_\bot)\nonumber
\end{equation}
\begin{equation}
\cdot J_{m^\prime} (k_\bot^\prime R_\bot)\sum_{n,n^\prime=0}^m
\left({\begin{array}{*{20}c}
   m  \\n\end{array}}\right)
   \left({\begin{array}{*{20}c}
   m^\prime  \\n^\prime\end{array}}\right)
\Big(-\frac{\mu}{M_e}\frac{r_\bot}{R_\bot}\Big)^{(n+n^\prime)}
e^{i(m+m^\prime-n-n^\prime)\varphi_R}e^{i(n+n^\prime)\varphi_r}.\label{eq:centered2}
\end{equation}
in the expression for the  relevant electromagnetic modes. As a
consequence, unless the atom is located on the beam axis, it
should be expected that the most important contributions to the
transition probabilities come from the $n=n^\prime=0$ terms in
these series. The two photons are then emitted producing a
translational and rotational recoil effect on the center of mass,
while the internal state of the atom remains invariant. Higher
order effects can be directly calculated using
Eqs.(\ref{eq:multipolar}-\ref{eq:multipolar2}).

\subsection{Matrix elements of the interaction term $H_{I3}$}

The interaction term $\hat H_{I3}$ makes it possible to change the
spin of the particles. The corresponding matrix elements can be
easily calculated within first order perturbation theory using the
identities
\begin{equation}
g_i\frac{q_i}{2M_i}{\bf S}_i\cdot \hat{\bf B}^{(TM)}({\bf r}_i) =
g_i\frac{q_i\omega}{4M_ick_z}E_0\Big[\psi_{m-1}({\bf r}_i)\hat
S_+^{(i)}+\psi_{m+1}({\bf r}_i)\hat S_-^{(i)}\Big]
\end{equation}
\begin{equation}
g_i\frac{q_i}{2M_i}{\bf S}_i\cdot \hat{\bf B}^{(TE)}({\bf r}_i) =
g_i\frac{iq_i}{4M_i}E_0\Big[\psi_{m-1}({\bf r}_i)\hat
S_+^{(i)}-\psi_{m+1}({\bf r}_i)\hat
S_-^{(i)}-\frac{2ik_\bot}{k_z}\psi_m({\bf r}_i)\hat
S_z^{(i)}\Big],
\end{equation}
with $S_{\pm}$ the ascending and descending spin operators. Again,
Eqs.~(\ref{eq:multipolar}-\ref{eq:multipolar2}) lead to a
multipole expansion for the matrix elements. For hydrogenic atoms,
the ratio $q_i/M_i$ is larger for electrons than for nuclei, so
that  the most probable event of this type produces changes of the
spin angular momentum of the electron by a factor $\pm \hbar$
without changing the  spatial wave function of relative motion,
while the center of mass acquires an angular momentum $-(m\mp
1)\hbar$. The event corresponding to no changes in the internal
wave function at the expense of a recoil effect with a change of
the orbital angular momentum of the center of mass by $-m\hbar$ is
also relevant for non paraxial photons. Notice that the matrix
elements of the center of mass calculated for the $H_{I1}$ can
also be used in the evaluation of the transition amplitudes
associated to $H_{I3}$.

\section{Comparison with previous results that use the paraxial approximation.}

Let us compare our results to those obtained by Babiker $et$ $al$
\cite{babiker4} who studied the transition amplitude for the
emission of electromagnetic photons of the generic type
\begin{equation}
{\bf A} ({\bf x},t) = \hat{\bf
\epsilon}F(x_\bot)e^{i(k_zx_z-\omega
t)}e^{im\varphi}\label{eq:generic}
\end{equation}
by hydrogenic atoms within the PZW formalism \cite{PZW,cohen}. The
fields (\ref{eq:generic}) can be considered a paraxial
approximation to the so called left and right polarized Bessel
modes \cite{barnett94,barnett2}
\begin{eqnarray}
{\bf A}^{({\cal L})}_m({\bf x},t;k_\bot,k_z) &=& A_0^{({\cal
L})}\Big[({\bf e}_x +i{\bf e}_y)\psi_m -
i\Big(\frac{k_\bot}{k_z}\Big)\psi_{m+1}{\bf e_z}\Big],\\
{\bf A}^{({\cal R})}_m({\bf x},t;k_\bot,k_z) &=& A_0^{({\cal
R})}\Big[({\bf e}_x -i{\bf e}_y)\psi_m +
i\Big(\frac{k_\bot}{k_z}\Big)\psi_{m-1}{\bf e_z}\Big].
\label{eq:circpol}
\end{eqnarray}
Their superpositions ${\bf A}^{({\cal R})}_m\pm{\bf A}^{({\cal
L})}_m$ define linearly polarized modes, and are linear
combinations of the elementary TE and TM modes:
\begin{eqnarray}
{\bf A}^{({\cal L})}_m &=&A_0^{({\cal L})\prime}\Big({\bf
A}^{(TM)}_{m+1}-i\frac{ck_z}{\omega}{\bf A}^{(TE)}_{m+1}\Big)
\label{eq:L}\\
{\bf A}^{({\cal R})}_m &=& A_0^{({\cal R})\prime}\Big({\bf
A}^{(TM)}_{m-1}+i\frac{ck_z}{\omega}{\bf
A}^{(TE)}_{m-1}\Big)\label{eq:R}.
\end{eqnarray}
Notice that an index $m$ for the polarized modes corresponds to
the superposition of $TE$ and $TM$ modes with $m\pm 1$. According
to the results we have obtained, the corresponding left and right
polarized Bessel beams carry an angular momentum $(m\pm 1)\hbar$
along the $z$ axis.

Having an explicit form for the generic function $F(x_\bot)$, we
obtained a complete multipole expansion that takes into account
both the azimuthal and radial behavior  of electromagnetic Bessel
modes. Accordingly, this transition amplitudes extend the results
of Babiker $et$ $al$  beyond the paraxial limit. Using
Eqs.~(\ref{eq:L}-\ref{eq:R}), it is straightforward to show that
our results are consistent with those reported in
Ref.\cite{babiker4}. However, there are some conceptual
differences between both approaches. The total angular momentum
for $TE$ and $TM$ photons is $m \hbar$ and cannot be directly
separated into 'orbital' or 'spin' parts. For the superpositions
of the $TE$ and $TM$ modes leading to ${\cal L}$ and ${\cal R}$
modes, this separation seems more natural and permited Babiker
$et$ $al$ to conclude that ``in the interaction of molecules with
light endowed with {\bf orbital} angular momentum, an exchange of
{\bf orbital} angular momentum in an electric dipole transition
occurs only between the light and the center of mass motion".

Nevertheless, the circularly polarized modes ${\cal R}$ and ${\cal
L}$ do not form an orthogonal basis. This has consequences when
performing a quantization of the electromagnetic field in terms of
them. By choosing $A_0^{({\cal L})\prime}=A_0^{({\cal R})\prime}
=\sqrt{1 +c^2k_z^2/\omega^2}/2$ the corresponding creation and
annihilation operators satisfy the standard commutation relations
\begin{eqnarray}
\big[\hat a^{({\cal L})}_m(k_z,k_\bot), \hat a^{({\cal
L})\dagger}_{m^\prime} (k_z^\prime,k_\bot^\prime)\big]
&=&\frac{1}{k_\bot}\delta_{m,m^\prime}
\delta(k_\bot-k_\bot^\prime)\delta(k_z-k_z^\prime)\nonumber \\
\big[\hat a^{({\cal R})}_m(k_z,k_\bot), \hat a^{({\cal
R})\dagger}_{m^\prime} (k_z^\prime,k_\bot^\prime)\big]
&=&\frac{1}{k_\bot}\delta_{m,m^\prime}
\delta(k_\bot-k_\bot^\prime)\delta(k_z-k_z^\prime).
\end{eqnarray}
However not all the other commutators are zero, in fact
\begin{equation}
\big[\hat a^{({\cal L})}_m(k_z,k_\bot), \hat a^{({\cal
R})\dagger}_{m^\prime+2} (k_z^\prime,k_\bot^\prime)\big]
=\frac{1-c^2k_z^2/\omega^2}{1+c^2k_z^2/\omega^2}\frac{1}{k_\bot}\delta_{m,m^\prime}
\delta(k_\bot-k_\bot^\prime)\delta(k_z-k_z^\prime).
\end{equation}
This results make it more difficult to interpret  the dynamical
observables of the field written in terms of creation and
annihilation operators. For instance, the energy is not diagonal
in this basis \cite{hacyan}.

\section{Conclusions and discussion.}
In this article we have calculated the emission probability
amplitude  of a Bessel photon by an atomic system within first
order perturbation theory. We obtained a closed expression of the
electromagnetic potentials that permits a systematic multipole
approximation taking into account both the azimuthal and radial
behavior  of the electromagnetic Bessel modes.

It was shown that the emission of a Bessel TE or TM photon of
order $m$ induces a change in the projection of angular momentum
along the $z$ axis of the atom that is always of magnitude $\hbar
m$. Thus, the angular momentum carried by these Bessel photons is
precisely $\hbar m$. This result is important because a field
theoretical description of the angular momentum for the
electromagnetic field in terms of $TE$ and $TM$ Bessel modes leads
to Eq.~(\ref{eq:angmom}). The second term in this equation is
usually related to spin angular momentum, while the surface
integral is not well defined. Thus, the calculations here
performed are a direct evaluation of the total angular momentum
$z$ component of a Bessel photon.

It is important to notice that the vectorial character of the
electromagnetic field is responsible for possible changes $\pm
\hbar$ in the internal angular momentum within the dipole
approximation. As usual, these changes are induced by the field
components along the circular vectors $\hat{\bf e}_x \pm \hat{\bf
e}_y$. The corresponding transition probabilities are proportional
to $E_0^2\sim k_z^2k_\bot c^2/\omega$ as can be seen from equation
(\ref{norm}). The transition probability of emission of $TM$
Bessel photons via $H_{I1}$ without changes in the internal
angular momentum necessarily leads to the maximum possible
exchange of angular momentum between a Bessel photon and the
center of mass motion. These transition probabilities are
proportional to $E_0^2k_\bot^2/k_z^2\sim k_\bot^3 c^2/\omega$ and
they could be important for non paraxial Bessel photons.

The fact that spontaneously emitted optical Bessel photons have
not been observed can be due to the small value of the center of
mass matrix elements under usual circumstances. In this article,
we have evaluated these elements both for free atoms and for
harmonically trapped atoms. In the first case, the conservation of
linear momentum in the radial direction do not single out a
particular value of $k_\bot^{R \prime}$; this is an important
difference with the result obtained in the axial direction for
which the matrix element is proportional to $\delta(k_z^{R\prime}
-k_z^{R}+k_z)$. Notice that both the radial and axial photon
functions are normalized via delta distributions; that is, the
matrix elements for the emission of a photon with specific
$k_\bot$, starting from an specified wave function of the center
of mass with a given $k_\bot^R$, are finite and different from
zero for a continuum range of values of the final transversal
momentum for the center of mass $k_\bot^{R\prime}$. As a
consequence, the idealized case of a transition between initial
and final states for the center of mass  represented by Bessel
functions, has an effectively zero probability. In any case,
comparison with experimental results would require working with
wave packets. For trapped atoms, the transverse part of the center
of mass wave function is localized. The transition amplitude
depends on the average position of the center of mass and on the
spread of the oscillation $\alpha$. We have shown that the matrix
element of the center of mass motion given by the standard
electric dipole matrix for the internal motion,
Eq.~(\ref{eq:dipole}), decays exponentially with the factor
$k_\bot^2\alpha^2/4$ relating the spread of the atom oscillation
to the photon wavelength. For paraxial beams and highly localized
trapped atoms, the condition $k_\bot^2\alpha^2 \ll 1$ is currently
feasible.

We have focused our attention on spontaneous transition
amplitudes, but  induced transition probabilities can also be
calculated from them using Einstein relationships. They will be
proportional to incident radiation intensity and could
experimentally confirm our results. We have explicitly shown that
certain mechanisms can enhance the probability of changing the
internal angular momentum of the atom in multiples of $\hbar$
larger than those predicted by the standard plane wave multipole
expansion. For instance, transitions with $\Delta m^r=\pm 2\hbar$
depend on the electric quadrupole matrix elements of the relative
motion and are given by {\bf two} kind of transition amplitudes:
one for the standard quadrupole expansion that is proportional to
$k_\bot$ and the other arising from terms with $n=1$ and $k=0$ in
Eq.~(\ref{eq:multipolar}). The latter are due to the vortex of the
Bessel mode in the beam axes and could be detected for a trapped
atom with an adequate value of $k_\bot^2\alpha^2/4$,
Eq.~(\ref{eq:centered2}).

Finally, we have analyzed some specific features of the transition
amplitudes associated to the interaction Hamiltonian $H_{I2}$ and
$H_{I3}$. In the long wavelength limit, the most important
transitions associated to the former Hamiltonian  involve two
photons that do not alter the internal motion of the atom but
exchange  linear, $\hbar (k_z + k^\prime_z)$, and angular
momentum, $(m+m^\prime)\hbar$, with the center of mass. For
hydrogenic atoms, the magnetic interaction $g_iq_i/2M_i{\bf
S}_i\cdot \hat{\bf B}({\bf r}_i)$ will favor changes in the spin
orientation of the electron $\pm \hbar$ and in the orbital angular
momentum of the center of mass.

\section*{Acknowledgements}
We acknowledge very stimulating discussions with Shahen Hacyan and
Karen Volke-Sep\'ulveda. This work was partially supported by
CONACyT 41048-A1.

\end{document}